# Effects of Carbon Nanotubes on Grain Boundary Sliding in Zirconia Polycrystals

Maren Daraktchiev[*], Bertrand Van de Moortèle, Robert Schaller, Edina Couteau, László Forró

Ecole Polytechnique Fédérale de Lausanne, Institut de Physique de la Matière Complexe,

CH-1015 Lausanne, Switzerland

Mechanical properties of zirconia polycrystals decrease drastically at high temperature due to thermally activated grain boundary (GB) sliding, leading to plastic or even super-plastic deformation.[1-5] As GB sliding is a source of energy dissipation in the material, mechanical loss measurements[6,7] are well suited to study such a mechanism. They reveal, in general, a mechanical loss peak, which evolves into an exponential increase at higher temperature.[8-11] When intergranular glassy films or/and amorphous pockets are presented in polycrystalline ceramics, the mechanical loss[12,13] is globally higher and so is the creep rate.[5,14,15] Here we show that introducing carbon nanotubes (CNTs) in zirconia, in particular, reduces drastically GB sliding and consequently the mechanical loss at high temperature. The nanotubes were observed at the grain boundaries by high-resolution transmission electron microscopy (HRTEM) and were related to the reduction of superplasic flow through the boundaries, which should improve the material creep resistance.

The mechanical loss, tan($\phi$), spectrum associated with the brittle-to-ductile transition in fully tetragonal zirconia polycrystals stabilized by 3 mol-% yttria (3Y-TZP) is presented in Figure1. As measured as a function of temperature, the mechanical loss angle tan($\phi$) shows an exponential increase accompanied by a steep shear modulus (G) decrease above 1200 K.

High temperature plastic deformation of polycrystalline zirconia is associated with GB sliding, and consequently it seems reasonable to link the mechanical loss spectrum with GBs.[16-18] A theoretical model[16] was developed to interpret the high temperature mechanical loss of Y-TZP as due to the relative sliding of grains of hexagonal shape separated by an intergranular viscous layer. From the view point of anelasticity, two forces play a role in this process: a dissipative force corresponding to the viscosity of the intergranular layer and a restoring force corresponding to the elasticity of the adjacent grains, which limit the sliding at the triple point junctions. Such a model accounts for a Debye peak.[7] However in the obtained spectrum (Fig. 1) a peak is not observed but an exponential increase in damping with temperature occurs. This behavior can be interpreted by considering that the restoring force due to elastic interaction at the triple points decreases when temperature rises. If the restoring force goes to zero there is no limitation to strain, and as a consequence, mechanical loss increases continuously with temperature. The signature of creep in the mechanical loss spectrum is an exponential form of damping (often called exponential background). On the other hand a relaxation peak is representative of an anelastic behavior, in which a restoring force limits strain. In the case of GB sliding in ceramics, a well-resolved peak would be observed only if the pinning centers, like GB asperities or triple point junctions, are strong obstacles. In the case of Y-TZP zirconia, the regular shape of the grains does not provide strong pinning centers at the triple points.

---

[*] Present Address: Department of Earth Sciences, University of Cambridge, Downung Street, Cambridge CB2 3EQ, United Kingdom; E-mail address: mdar04@esc.cam.ac.uk

The goal of the present research was therefore to improve the high-temperature (micro)creep resistance of 3Y-TZP by reinforcing the microstructure with multiwalled carbon-nanotubes (MWCNTs). On the one hand, carbon nano-objects can occupy the intergranular spaces in 3Y-TZP (see Fig.2, 3). On the other hand, as CNTs are objects with low density of structural defects and covalent inter-atomic bonds[19], they could improve the mechanical properties of 3Y-TZP ceramics. An attractive example of this issue in ceramics is Zhan's investigations[20], which have demonstrated that the fracture toughness in CNTs-reinforced $Al_2O_3$ composites is three times as high as alumina materials. However these authors did not investigate the effect of the CNTs on the high temperature mechanical properties, which is primordial for applications. In the present paper we show that CNTs improve dramatically the high temperature strength of 3Y-TZP. For comparison, we also measured 3Y-TZP doped with 5 wt.-% of $SiO_2$ amorphous phases. $SiO_2$ inclusions present a lubricating agent for grain-boundary sliding in zirconia ceramics, which enhances their plastic strain. In summary, three grades of 3Y-TZP were processed by cold pressing (130 kN) and sintering at 1725 K for 180 min.: pure 3Y-TZP; $SiO_2$ – doped 3Y-TZP; and CNTs-doped 3Y-TZP (see Experimental). The mechanical loss measurements were carried out in an inverted torsion pendulum working with forced vibrations.

The results obtained in 3Y-TZP, $SiO_2$ doped and CNTs reinforced 3Y-TZP are presented in Fig.4. As already seen in Fig.1, the mechanical loss of 3Y-TZP ceramics shows an exponential increase as a function of temperature associated with a rapid decrease in material stiffness. No dissipative peak is observed in 3Y-TZP between 1200 and 1600 K, which shows that the anelastic strain has no limit in this temperature range. Doping of 3Y-TZP with $SiO_2$ lubricating phases results in an increase of $\tan(\phi)$ by a factor of 2 with respect to pure 3Y-TZP (Fig.4), and a new relaxation peak appears at 1400 K.

As it concerns GB sliding, it is possible to conclude that the higher the amount of lubricating agent, the easier the grain boundary slip and the higher the energy dissipation. The peak interpretation depends on the distribution of the $SiO_2$ phases in the 3Y-TZP structure at grain boundaries or at multiple points. Intergranular $SiO_2$ phase acts as a shortcut for matter transport along the grain boundaries.[21] Thus, the peak in 3YTZP+$SiO_2$ samples, having a preferential $SiO_2$ distribution at grain boundaries, can be associated with an acceleration of grain-boundary sliding kinetics in these samples. The distribution of $SiO_2$ additions as forming glassy pockets at multiple points may also account for a dissipative peak in $SiO_2$ doped 3Y-TZP, due to the vitreous transition (the $\alpha$-relaxation) in the amorphous pockets. Similar situation arises in $Si_3N_4$ composites, for example, where a mechanical loss peak results from the amorphous pockets, the surrounding skeleton being responsible for creep resistance[12] or in bulk metallic glasses around the temperature of glass transition.[22]

The mechanical loss spectrum of "3YTZP+CNTs" shows a relaxation peak at 1500 K, which is superimposed on an exponential background at higher temperature (see Fig.4). The comparison of the mechanical loss in CNTs-3Y-TZP composite with pure 3Y-TZP and $SiO_2$ doped 3Y-TZP shows a significant decrease in tan($\phi$) due to the CNTs inclusions. Note also that tan($\phi$) in CNTs-3Y-TZP is displaced towards higher temperature, which indicates that these composites may have a better creep resistance. TEM observations in CNT-3Y-TZP sample reveal a microstructure similar to the "traditional" Y-TZP with a grain size around 0.1 ÷ 0.5 μm.[23] The high diffraction contrast along the grain boundaries is due to large residual stresses, resulting from the incorporation of CNTs at grain boundaries (see Fig.2). An HRTEM image of such an area is reported in Fig.3a. Two zirconia grains (top and bottom in Fig.3a) are separated by a very thick amorphous layer. This layer is not consistent with the intergranular glassy-phase thickness (about 0.8 nm), being calculated and observed in different ceramic systems.[15,24] Figures 3b,c correspond to a qualitative electron energy loss spectroscopy (EELS) analysis performed in both the zirconia area and the intergranular layer. In Fig.3b, the Zr $M_{45}$ and $M_2$ edges (start at 180 eV and 344 eV, respectively) are clearly visible. The edge at 532 eV corresponds to O-edge. In Fig. 3c, the most important edge is the C edge (starts at 280 eV), which indicates that the amorphous layer between the zirconia grains corresponds to amorphous carbon. The Zr and O edges are still present in Fig. 3c. Their existences could be explained with either the difficulty to analyze very small areas of grain boundaries without destroying the neighboring areas or the presence of zirconia grain(s) below the point of EELS analysis. The relaxation peak at 1500 K (see Fig.4) can be thus associated with the microstructural changes at grain boundaries or triple points, which appear after the CNTs reinforcement. By analogy with previous relaxation models,[6,7] the lower level of tan($\phi$) in 3Y-TZP+CNTs could be attributed to a decrease in GB slip via an interaction with the intergranular nanotubes inclusions (bigger restoring force).

Mechanical spectroscopy was used to investigate the mechanical properties of a new composite: 3Y-TZP reinforced with CNTs. The results show clearly that this composite material exhibits a better resistance to GB sliding at high temperature, which may lead to a better creep behavior. Moreover the CNTs amorphization in our samples has not been observed in other CNTs reinforced ceramics.[20] Zhan et al.[20] have shown CNTs network being developed along the alumina grains. Thus, the CNTs amorphization is probably due to our method of processing. Instead of spark-plasma sintering proposed elsewhere[20], we performed "conventional" sintering (at 1725 K, 180 min) on 3YTZP+CNTs powders, which might have destroyed the crystalline structure of carbon sheets (note that some crystalline domains still exist inside the amorphous layer in Fig.3a). However, the mechanical properties of multiwalled CNTs are not as good as single-walled CNTs (SWCNTs) Despite the use of MWCNTs, we obtained a 3YTZP+CNTs composite with a very good high-temperature behavior (see Fig.4). In our opinion, the mechanical properties of MWCNTs increase considerably into the 3Y-TZP matrix, because of hindering of the individual sheet sliding.

*Acknowledgments*

This work was supported by the Swiss National Science Foundation and in part by the TOP NANO 21 Program. The authors would like to thank L. Gremillard for his assistance in TEM observations, B. Guisolan for technical help and Thomas Lagrange for critically reading the manuscript. The CLYME (Consortium LYonnais de Microscopie Electronique) were grateful acknowledged for the access of this microscope.

*Experimental*

*Materials:* 3Y-TZP powder was supplied from Tosoh Corporation, Japan. The carbon nanotubes were produced by CVD on 5% Fe(III),Co/CaCO$_3$ as catalyst and purified by dissolution and sonication in diluted nitric acid for 30 minutes, followed by filtration, washing and drying at 393 K for about one night.[25] 3Y-TZP samples were prepared by cold pressing and sintering from 3Y-TZP powder. SiO$_2$ doped 3Y-TZP was processed via slip-casting method (5 wt% of colloidal silica "Ludox HS40" and 3Y-TZP) followed by cold pressing and sintering. CNTs doped 3Y-TZP was prepared by mixing of 3Y-TZP powder with 5 wt% of MWCNTs, followed by homogenizing of the mixture via attrition milling by using 20 mm zirconia ball grinding media, cold pressed and sintered). All samples were sintered at 1753 K for 180 minutes: high-purity 3Y-TZP and SiO$_2$ doped 3Y-TZP were sintered under air, while CNTs doped 3Y-TZP – in nitrogen atmosphere.

*Characterization:* Sample for TEM observations were prepared by typical dimpling polishing and ion beam milling. The beam voltage was about 3.5 keV and tilt angle between 5-6 degrees. A JEOL 2010F apparatus, with a field emission operating at 200 keV, was used to HRTEM observations. This microscope is equipped with a GATAN DigiPEELS spectrometer and an INCA analyzer for chemical measurements. The Conventional TEM was performed on a JEOL 200CX in the GEMPPM laboratory. Mechanical loss measurements[26] were carried out in a differential forced pendulum under vacuum of order of $10^{-3}$ Pa. Two flat bar specimens are mounted in series along the pendulum axis, one being measured (3Y-TZP samples) and the other one acting as an elastic reference (WC-6vol%Co). From sample angular deformations one estimates the mechanical loss angle tan($\phi$) and the normalized shear modulus $G/G_{300K}$ of the 3Y-TZP samples either as a function of temperature (300 – 1600 K) at fixed frequencies, or as a function of frequency ($10^{-3}$ – 10 Hz) at fixed temperatures.

**REFERENCES**

[1] F. Wakai, S. Sakaguchi, Y. Matsuno, *Adv. Ceram. Mater* **1986**, *1*, 259.

[2] F. Wakai, N. Kondo, H. Ogawa, T. Nagano, S. Tsurekawa, *Mater. Charact.* **1996**, *37*, 331.

[3] M. Jiménez-Melendo, A. Domínguez-Rodríguez, *Acta Mater.* **2000**, *48*, 3201.

[4] D. Owen, A. Chokshi, *Acta Mater.* **1998**, *46*, 667.

[5] M. Nauer, C. Carry, *Scripta Metall. Mater.* **1990**, *24*, 1459.

[6] A. Nowick, B. Berry, *Anelastic relaxation in crystalline solids*; Academic Press, New York **1972**.

[7] R. Schaller, G. Fantozzi, G. Gremaud, *Mechanical spectroscopy Q-1 2001 with applications to materials science*; Trans Tech Publications **2001**.

[8] R. Schaller, *J. Alloy. Compd.* **2000**, *310*, 7.

[9] S. Testu, R. Schaller, J. Besson, T. Rouxel, G. Bernard-Granger, *J. Eur. Ceram. Soc.* **2002**, *22*, 2511.

[10] L. Donzel, A. Lakki, R. Schaller, *Phil. Mag. A* **1997**, *76*, 933.

[11] G. Pezzotti, *J. Am. Ceram. Soc.* **2001**, *84*, 2225.

[12] A. Lakki, R. Schaller, G. Bernard-Granger, R. Duclos, *Acta Metall. Mater.* **1995**, *43*, 419.

[13] L. Donzel, E. Conforto, R. Schaller, *Acta Mater.* **2000**, *48*, 777.

[14] Y. Lin, P. Angelini, M. Mecartney, *J. Am. Ceram. Soc.* **1990**, *73*, 2728.

[15] K. Hiraga, H. Y. Yasuda, Y. Sakka, *Mater. Sci. Eng. A* **1997**, *234-236*, 1026.

[16] A. Lakki, R. Schaller, M. Nauer, C. Carry, *Acta metall.* **1993**, *41*, 2845.

[17] M. Daraktchiev, R. Schaller, *phys. stat. sol.(a)* **2002**, *195*, 293.

[18] R. Schaller, M. Daraktchiev, *J. Eur. Ceram. Soc.* **2002**, *22*, 2461.

[19] M. Dresselhaus, G. Dresselhaus, P. Avouris, *Carbon nanotubes: Synthesis, Structure, Properties, and Applications*, Spriger Verlag, Berlin **2001**

[20] G. Zhan, J. Kuntz, J. Wan, A. Mukherjee, *Nature Mater.* **2003**, *2*, 38.

[21] F. Wakai, *Acta Metall. Mater.* **1994**, *42*, 1163.

[22] J. Pelletier, B. V. D. Moortèle, I. Lu, *Mater. Sci. Eng. A* **2002**, *336*, 190.

[23] L. Gremillard, T. Epicier, J. Chevalier, G. Fantozzi, *Acta Mater.* **2000**, *48*, 4647.

[24] D. Clarke, *J. Am. Ceram. Soc.* **1987**, *70*, 15.

[25] E. Couteau, K. Hernadi, J. W. Seo, L. Thien-Nga, C. Miko, R. Gaal, L. Forro, *Chem. Phys. Lett.* **2003**, *378*, 9.

[26] M. Daraktchiev, *High temperature mechanical loss in yttria stabilized tetragonal zirconia and in calcium hexaluminate*; EPF, Lausanne, **2003**.

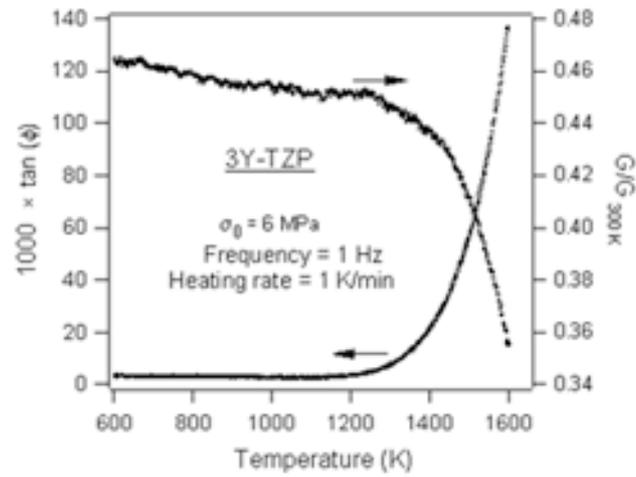

**Figure 1:** Mechanical loss, tan(ϕ), in 3Y-TZP as measured as a function of temperature for a frequency of 1 Hz and heating rate of 1 K/min. The exponential increase in tan(ϕ) above 1200 K is accompanied by a steep modulus decrease.

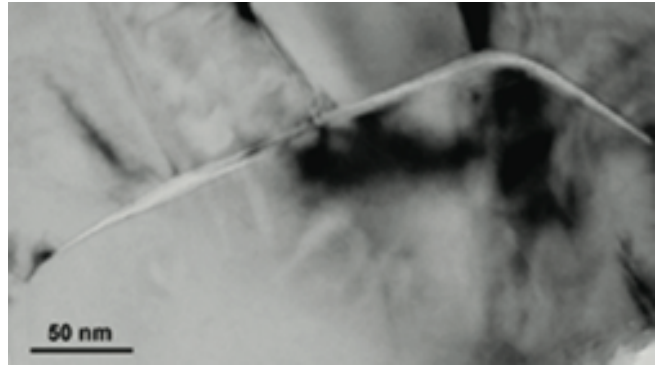

**Figure 2:** Bright Field image of 3Y-TZP reinforced MWCNT composite. Strong bright contrast along the grain boundaries could be observed in the centre of this image. The black regions are due to diffraction contrast from residual stresses in the grains.

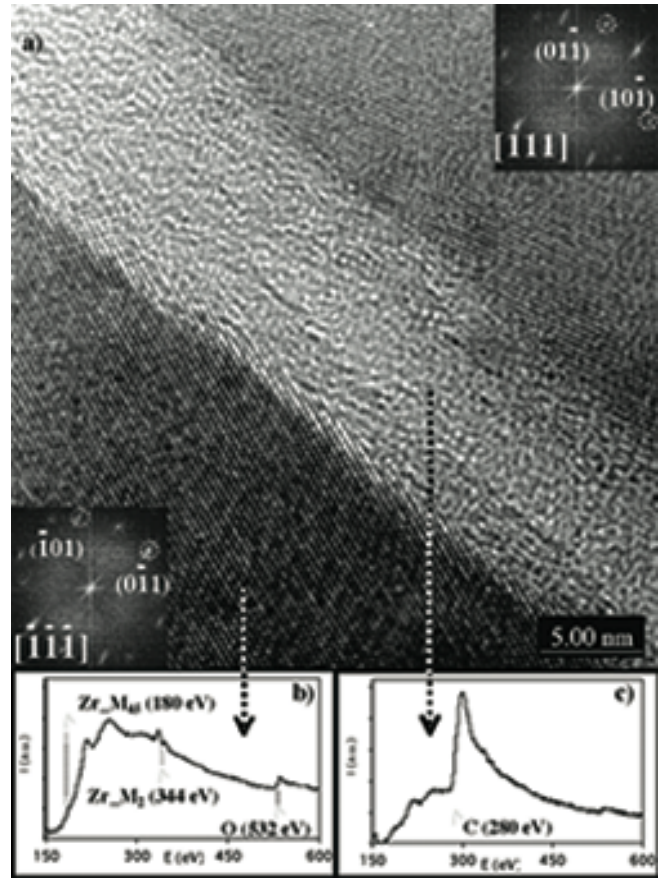

**Figure 3: a)** HRTEM image and Fourier-transform of the grain boundaries in 3Y-TZP reinforced with MWCNTs. Grains were identified as tetragonal phase[21] (a ≈ 0.3606 nm and c ≈ 0.5176 nm) with azimuth [1 1 1] and [$\bar{1}$ $\bar{1}$ $\bar{1}$] for the top and bottom grains, respectively; **b, c)** Electron energy loss spectroscopy (EELS) spectra of the grain boundaries. EELS spectra are obtained in different areas as indicated by the arrows.

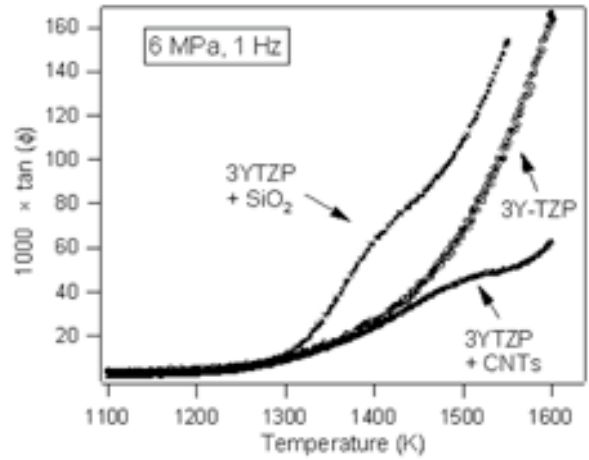

**Figure 4:** Mechanical loss spectra of 3Y-TZP, $SiO_2$ doped and carbon-nanotubes (CNTs) doped 3Y-TZP ceramics. The exponential increase in zirconia polycrystals is the highest in $SiO_2$ doped 3Y-TZP, because the silica phase enhances grain boundary sliding. The CNTs inclusions form additional pinning points that increase the restoring force and limit the grain displacements, which results in a decrease of mechanical loss.